\documentclass[aps,prb,twocolumn,groupedaddress,floats,showpacs]{revtex4-1}

\usepackage{latexsym}
\usepackage{dcolumn}
\usepackage[dvips]{graphicx}
\usepackage{amssymb}
\usepackage{graphics}
\usepackage{amsmath}
\usepackage{epsf}

\begin{document}
\title{Plasmon-phonon coupling in graphene}
%\author{E. H.\ Hwang, S.\ Das Sarma, R. Sansarma}
\author{E. H. Hwang, Rajdeep Sensarma, and S. Das Sarma}
\address{Condensed Matter Theory Center, Department of Physics 
	 University of Maryland 
	 College Park, Maryland  20742-4111}
\date{\today}

\begin{abstract}
Collective excitations of coupled 
electron-phonon systems 
are calculated for both monolayer and bilayer graphene, taking into
account the non-perturbative Coulomb coupling between electronic
excitations in graphene and the substrate longitudinal optical phonon modes.
We find that
the plasmon-phonon coupling in monolayer graphene is strong at all
densities, but in bilayer graphene the coupling is significant only at high
densities satisfying the resonant condition $\omega_{pl} \approx \omega_{ph}$. 
The difference arises 
from the peculiar screening properties associated with chirality
of graphene.  
Plasmon-phonon coupling explains the measured
quasi-linear plasmon  dispersion in the long wavelength limit, thus
resolving a puzzle in the experimental observations.
\end{abstract}


\maketitle

A plasmon is a collective mode of charge-density oscillation in the
free-carrier system, which is present both in classical and quantum
plasmas.  
%The plasma excitation is a fundamental elementary excitation
%of any charged-particle system.  
Studying the collective plasmon
excitation in the electron gas has been among the very first
theoretical quantum mechanical many-body problems studied in
solid-state physics.  The collective plasmon modes of monolayer
graphene (MLG) have
been extensively studied theoretically
\cite{hwang2007,hwang2009,dassarma2009,wunsch2006,wang2007,vafek2006}
and experimentally
\cite{kramberger2008,liu2008,liu2010,langer2010,lu2009}.
Recent
discovery of bilayer graphene (BLG)
has also led to a number 
of theoretical descriptions of plasmon modes in BLG
\cite{MacDonaldBLG2009,sensarma2010}.  Even 
though the long-wavelength plasmon frequency of MLG is explicitly
nonclassical (i.e., the plasmon frequency is necessarily
quantum with “$\hbar$” appearing manifestly in the long-wavelength
plasma frequency \cite{dassarma2009}), its wave vector dispersion is given by
classical electrodynamics, i.e., $\omega_p(q) \propto \sqrt{q}$. Note
that the
quadratic band-dispersion of BLG makes the leading order
long-wavelength plasmon dispersion explicitly classical with the same
$\sim \sqrt{q}$ dependence. However, at finite $q$, away from the
long-wavelength limit, there are several corrections to the plasmon
dispersion $\omega(q)$ arising from nonlocal finite wave vector
response, finite-temperature thermal corrections, many-body effects,
local-field corrections, and other mechanisms relevant to the specific
electron system.

Since the plasmon dispersion relation is exactly known at long
wavelengths ($q \rightarrow 0$) where the f-sum rule arising from
particle conservation fixes the plasma
frequency, it is surprising that the measured graphene plasmon
dispersion in the long wavelength limit deviates from the classical
dispersion ($\omega_q \sim \sqrt{q}$) and shows a rather linear dispersion
\cite{liu2008,langer2010}. In a recent experiment \cite{liu2010}, the
strongly coupled plasmon-phonon mode dispersion has been measured by the
angle-resolved reflection electron-energy-loss spectroscopy and it is
found that the discrepancy  arises from 
electron-phonon coupling.  In epitaxial graphene the
substrate (i.e. SiC) is a highly polar material.  In general, carriers in
polar materials couple with the longitudinal optical (LO) phonons of
the system via the long-range Fr\"{o}hlich interaction.
However, the 
surface optical (SO) phonon is a well-characterized surface property
of polar semiconductors, and it is possible that carriers in graphene
layer couple to the SO-phonons of the underlying substrate
lattice via the long-range polar Fr\"{o}hlich coupling
\cite{fuchs1965,wang1972}.  For isotropic media the frequency of SO
phonons $\omega_{SO}$ is related to the transverse optical (TO) bulk
phonon $\omega_{TO}$ as
%\begin{equation}
${\omega_{SO}}/{\omega_{TO}}=\sqrt{({\epsilon_0+1})/({\epsilon_{\infty}
    +1})}$,\cite{dubois1982,fuchs1965,wang1972} 
%\end{equation}
where $\epsilon_0$
($\epsilon_{\infty}$) is the static (high frequency) dielectric constant. 
Note that the bulk longitudinal optical phonons $\omega_{LO}$ 
and $\omega_{TO}$
are connected with the dielectric constants by the
Lyddane-Sachs-Teller relation
${\omega_{LO}}/{\omega_{TO}} = \sqrt{
  {\epsilon_0}/{\epsilon_{\infty}}}$.

The electron-phonon coupling is the macroscopic coupling of the
electronic collective modes (plasmons) to the optical phonons.  The
mode coupling phenomenon, which hybridizes the collective plasmon
modes of the electron gas with the optical-phonon modes of the
lattice, gives rise to the coupled plasmon-phonon modes (the hybrid
modes) which have been extensively studied
\cite{abstreiter1984,matz1981,jalabert1989,hwang1995}
both experimentally and theoretically in bulk and 2D electron
systems. The electron-phonon interaction leads to many-body
renormalization of the single-particle free carrier properties
\cite{tse2008,park2009} and also affects the transport
properties \cite{fratini2008,chen2008}.  A good understanding of
electron-phonon coupling is thus important in developing quantitative
theories for many different experimental studies in graphene.  

In this paper we calculate the coupled plasmon-SO phonon modes of
epitaxial graphene (or graphene on a polar substrate such as
SiO$_2$, SiC, or HfO$_2$).  Our most 
significant finding is that in MLG plasmon-phonon mode coupling
effect is strong at all electronic densities due to the
singular behavior in the screening function arising from
chirality \cite{hwang2007}.
%We also extend our analysis to epitaxial BLG
%treated within a two-band parabolic dispersion model. 
By contrast, for BLG, 
the plasmon-phonon coupling is significant only at 
high carrier densities. We also find
that at low densities, when the coupling is weak and the 
coupled phonon-like mode (gapped mode) lies in the interband
electron-hole continuum, the 
energy of phonon-like mode decreases in the long wavelength limit due
to the coupling of the phonon mode to the interband single
particle excitation, which arises from the
enhanced BLG backscattering~\cite{HwangBLG2008}.
However, at high densities, when the plasmon-phonon mode
coupling is strong, the phonon-like mode frequency increases linearly with
wavevector, as in MLG.

We first present our model for plasmon-phonon coupling,
% in MLG.
%The model can be extended in a fairly straightforward way to
%bilayer systems and we will indicate the points of departure as we
%develop the model. 
which consists of a two-dimensional electron
gas coupled to dispersionless SO-phonons at zero temperature. For
MLG, we have a system of Dirac fermions with linear
dispersion, while BLG have a parabolic dispersion around
the Dirac point. Due to the presence of the long range electron-phonon
coupling, electrons interact among themselves through the Coulomb
interaction and through virtual-SO-phonon exchange via the
Fr\"{o}hlich interaction.  The electron-SO phonon interaction is given
by
\begin{equation}
H_{e-ph}=\sum_{kq}\sum_{ss'}M_{kq}^{ss'}c_{k+qs'}^{\dagger}c_{ks} \left (
b_q + b_{-q}^{\dagger} \right ),
\end{equation}
where $c_{ks}^{\dagger}$ is the electron ($s=+1$) or hole ($s=-1$)
creation operator, $b_q^{\dagger}$ and $b_q$ are creation and
destruction operators of surface phonon, and the interaction matrix
element $M_s^{ss'}$ is defined by
\begin{equation}
M_{kq}^{ss'}=M_0(q)F_{sk+q}^{\dagger}F_{s'k},
\end{equation}
where $F_{sk}$ is the chiral spinor and given by 
$F_{sk}^{\dagger}= (s, e^{-i\theta_k})/\sqrt{2}$
with $s=\pm1$, $\theta_k=\tan^{-1}(k_y/k_x)$ for MLG~\cite{ando2006}
and $\theta_k=2\tan^{-1}(k_y/k_x)$ for BLG~\cite{AndoBLG2006,mccann2006}. We also have 
\begin{equation}
[M_0(q)]^2 = \frac{2\pi e^2}{q}e^{-2qd}\frac{\omega_{SO}}{2}\left [
  \frac{1}{\epsilon_{\infty}+1} - \frac{1}{\epsilon_0+1} \right ],
\end{equation}
where $d$ is the separation distance between graphene layer and substrate.
Then the matrix element of SO-phonon
mediated electron-electron interaction is dependent on both wave vector and
frequency and give by, 
\begin{equation}
v_{ph}(q,\omega)=[M_0(q)]^2D_0(\omega),
\end{equation}
where the unperturbed SO-phonon propagator is given by
$D_0(\omega)={2\omega_{SO}}/({\omega^2-\omega_{SO}^2})$.

%%%%%%%%%%%%%%%%%%%Fig 1%%%%%%%%%%%%%%%%
\begin{figure}
\includegraphics[width=8.5cm]{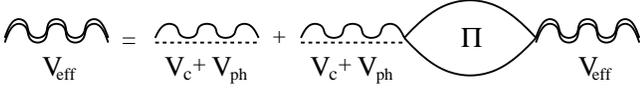} 
\caption{Effective dynamical interaction $V_{eff}$ calculated in
  RPA. Dashed (wiggly) lines represent the SO-phonon mediated
  (Coulomb) electron electron
  interaction $V_{ph}$ ($v_c$), and the bubble the irreducible polarizability
  $\Pi(q,\omega)$.
\label{diagram}
}
\end{figure}

The total effective electron-electron interaction is obtained in
RPA \cite{jalabert1989} by summing all the bare bubble diagrams (see
Fig.~\ref{diagram}), 
\begin{equation}
v_{\rm eff}(q,\omega) = \frac{v_c(q) + v_{ph}(q,\omega)}{1 - \left [ v_c(q)
+ v_{ph}(q,\omega) \right ] \Pi_0(q,\omega)} =  \frac{v_c(q)}{\epsilon_t(q,\omega)}
\end{equation}
where $v_c(q)=2\pi e^2/\epsilon_{\infty} q$ is the electron-electron
Coulomb interaction and $\Pi_0(q,\omega)$ is the complex irreducible
polarizability of either the monolayer\cite{hwang2007} or the
bilayer system~\cite{sensarma2010} given by the bare bubble diagram.  The
total dielectric function within RPA contains contributions both from
electrons and SO-phonons:
\begin{equation}
\epsilon_t(q,\omega) =  1- \frac{2 \pi e^2}{\epsilon_{\infty}q}
\Pi_0(q,\omega) + 
\frac{\alpha e^{-2qd}}{1 - \alpha e^{-2qd} -
\omega^2/\omega_{SO}^2},
\end{equation}
where 
\begin{equation}
\alpha = \epsilon_{\infty} \left [ \frac{1}{\epsilon_{\infty}+1} -
  \frac{1}{\epsilon_0 +1} \right ].
\end{equation}
The collective mode dispersion is given by the zeros of the
complex total dielectric function: $\epsilon_t(q,\omega)=0$.

Let us first focus on the collective modes of MLG. In the
long wavelength limit ($q\rightarrow 0$) we get the following coupled
$\omega_{\pm}$ collective modes
\begin{subequations}
\begin{eqnarray}
\omega_+(q) & = & \omega_{SO} + \frac{\alpha e^{-2qd}}{\omega_{SO}}
\frac{\omega_q^2 }{2}, \\
\omega_-(q) & = & \left ( 1- \alpha e^{-2qd} \right ) \omega_q,
\end{eqnarray}
\end{subequations}
where $\omega_q^2 =  2 e^2 E_F q/\epsilon_{\infty}$ ($E_F=$ Fermi
energy) is the plasmon
mode dispersion of an uncoupled system in the long wave length
limit. As $q \rightarrow 0$ the phonon-like mode $\omega_{+}$ is located above
$\omega_{SO}$ and increases linearly, and the plasmon-like
$\omega_{-}$ is slightly less than the 
corresponding uncoupled monolayer graphene plasmon mode,
$\omega_q$. 

\begin{figure}
%\epsfysize=2.in
%\epsffile{c_mode.eps}
%\epsfysize=2.in
%\epsffile{c_mode_l.eps}
\includegraphics[width=\columnwidth]{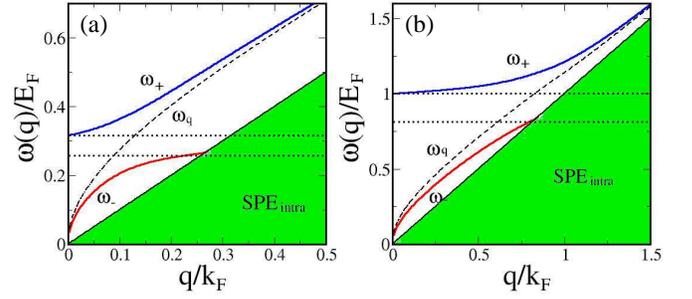} 
\caption{Calculated plasmon-phonon coupled mode $\omega_{\pm}$
  dispersions in MLG as a 
  function of the wave vector $q$ for two different densities (a)
  $n=10^{13} cm^{-2}$ and (b) $n=10^{12} cm^{-2}$. The plasmon dispersion
  ($\omega_q$)  without the electron-phonon coupling is shown
  by the dashed line. Two dotted horizontal lines represent the
  frequencies of the uncoupled SO (top) and TO (bottom) phonon modes,
  respectively. 
\label{fig_slg_mode}
}
\end{figure}

In Fig.~\ref{fig_slg_mode} we show the calculated coupled
plasmon-phonon collective modes in MLG for two
different densities. 
%We show the coupled modes ($\omega_{\pm}$) as
%well as the uncoupled plasmon mode ($\omega_q$).  
The following
parameters are used throughout this paper \cite{nienhaus1995}:
$\omega_{TO}=95.0$ meV, $\omega_{SO}=116.7$ meV, $\epsilon_{\infty} =
6.4$, $\epsilon_0=10.0$, and $d=5\AA$.  As shown in
Fig.~\ref{fig_slg_mode} the mode coupling in MLG is strong for
all electron densities.  In ordinary 2D systems or 3D systems the
plasmon-phonon mode coupling is only significant at densities
satisfying the resonant condition $\omega_{q} \approx
\omega_{SO}$. However, in MLG the plasmon mode exists
for all wave vectors due to the singular behavior in the
polarizability, which leads to strong plasmon-phonon coupling.  Since
the singular behavior of the polarizability arise from the suppression
of the back scattering due to the chirality of MLG the strong
plasmon-phonon coupling is a direct consequence of its unique chiral
property of MLG.  Note that the plasmon-like mode $\omega_{-}$ in
Fig.~\ref{fig_slg_mode} vanishes at a finite critical wave vector, $q_c
\simeq \omega_{SO}(1-\alpha)/v_F$, and for $q>q_c$ we find only the
phonon-like mode ($\omega_{+}$) which approaches $\omega_q$ for large
$q$.

\begin{figure}
\includegraphics[width=\columnwidth]{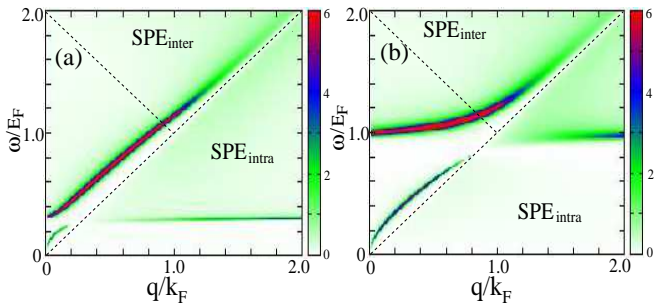} 
\caption{The density plots of energy loss function ($-{\rm
    Im}[1/\epsilon(q,\omega)]$) of MLG in ($q,\omega$) space 
  for two different densities (a)
  $n=10^{13} cm^{-2}$ and (b) $n=10^{12} cm^{-2}$. 
Note that the $\omega_+$ mode of the high density result carries most
spectral weight and its dispersion is almost linear, which is observed
in the recent experiment \cite{liu2008}. We use a phenomenological
damping of $0.1E_F$ in these results.
}
\label{fig_spec}
\end{figure}

The dynamical structure factor, $S(q,\omega)$, which gives the spectral weight
of the collective modes, is proportional to the imaginary
part of the inverse dielectric function (loss function) and given by
\begin{equation}
S(q,\omega)=-\frac{1}{n_0 v_c(q)} {\rm Im} \left [
\frac{1}{\epsilon_t(q,\omega)} \right ].
\end{equation}
For a true collective mode with zero Landau damping both
Im[$\epsilon_t(q,\omega)$] and Re[$\epsilon_t(q,\omega)$] vanish, and
the inverse dielectric function becomes a delta function with weight
\begin{equation}
W(q)=\frac{\pi}{\frac{\partial}{\partial \omega} {\rm Re}
\epsilon_t(q,\omega)|_{\omega = \omega_i(q)} } 
\end{equation}
where $\omega_i(q)$ is the collective mode
frequency at wave vector $q$.
In the long wavelength limit the weight of plasmon-like
mode can be calculated as 
\begin{equation}
W(q)|_{\omega_-}=\frac{\pi}{2} (1-\alpha)^{3/2}\omega_q
\end{equation}
and the weight of phonon-like mode as 
\begin{equation}
W(q)|_{\omega_+}={\pi} \alpha \omega_{SO}/2.
\end{equation}
The spectral weight of $\omega_-$ mode vanishes as $\sqrt{q}$ in the
long wavelength limit, but 
the weight of $\omega_+$ mode is finite. Thus in the long wavelength
limit all spectral weight is carried by the phonon-like mode.
In Fig.~\ref{fig_spec} the calculated loss function $-{\rm
  Im}[1/\epsilon(q,\omega)]$  is shown in ($q,\omega$) space
for two different  densities (a) $n=10^{13} cm^{-2}$ and (b) $n=10^{12}
cm^{-2}$. 
In the long wavelength limit the phonon-like mode has most of the
weight. In the intermediate wave vector range,
however, the plasmon-like mode becomes stronger. 
The weight of the $\omega_{-}$ mode vanishes again when
the plasmon-like mode merges with the 
electron-hole continuum at a critical wave vector and $\omega_-$ mode
becomes overdamped by Landau damping.
\begin{figure}
%\epsfysize=2.in
%\epsffile{c_mode.eps}
%\epsfysize=2.in
%\epsffile{c_mode_l.eps}
\includegraphics[width=\columnwidth]{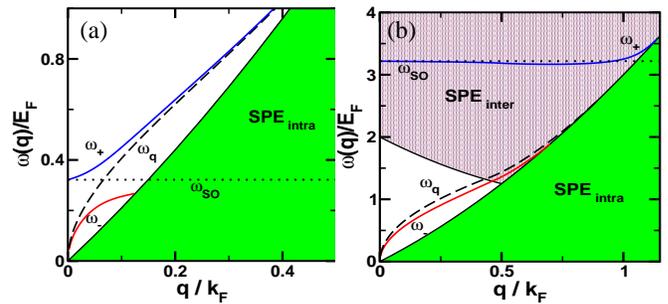} 
\caption{Calculated plasmon-phonon coupled mode $\omega_{\pm}$
  dispersions in bilayer graphene as a 
  function of the wave vector $q$ for two different densities (a)
  $n=10^{13} cm^{-2}$ and (b) $n=10^{12} cm^{-2}$. The plasmon dispersion
  ($\omega_q$)  without the electron-phonon coupling is shown
  by the dashed line. The dotted horizontal line represent the
  frequency of the uncoupled SO phonon mode.
\label{fig_blg_mode}
}
\end{figure}

Let us now turn our attention to BLG. Just like
MLG, we again get two hybridized plasmon-phonon
modes, one ($\omega_-$) having a $\sim \sqrt{q}$ dispersion and the
other ($\omega_+$) exhibiting a gap equal to the SO phonon frequency
($\omega_{SO}$) in the long wavelength limit. The $\omega_-$ mode has
the same dispersion as in MLG,
$\omega_-(q)=(1-\alpha e^{-2qd})\omega_q$, which lies in the gap
between the intraband and interband continua and has a spectral
weight which goes as $\sim \sqrt{q}$. Thus, in the long wavelength
limit, all the oscillator strength lies in the gapped mode
$\omega_+$.

\begin{figure}[t]
\includegraphics[width=\columnwidth]{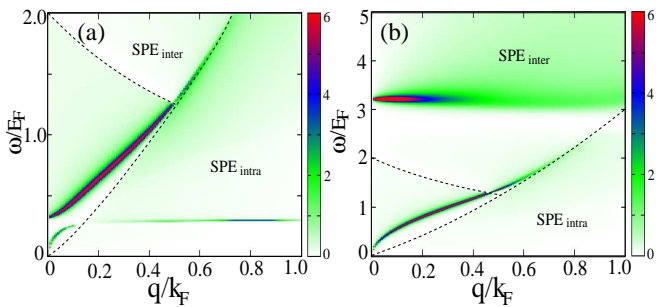} 
\caption{The density plots of energy loss function ($-{\rm
    Im}[1/\epsilon(q,\omega)]$) of bilayer graphene in ($q,\omega$)
  space for two different densities (a) $n=10^{13} cm^{-2}$ and (b)
  $n=10^{12} cm^{-2}$. }
\label{fig_spec_blg}
\end{figure}

However, there are two main differences from MLG,
i.e., the quadratic energy dispersion and the enhanced backscattering due to
chirality in BLG~\cite{HwangBLG2008},
which lead to non-trivial differences in the collective mode spectrum.
%i) The quadratic as opposed to linear dispersion of the single
%particle excitations
% which lead to the density dependence of the
%dimensionless ratio of kinetic to interaction energy scales, $r_s$,
%and ii) the doubling of the angle in the chiral spinors, which lead to
%enhanced rather than supressed backscattering in
%BLG~\cite{HwangBLG2008}. 
These two effects lead to very 
different behaviour in the low and high density limits. 
To illustrate
these effects, we plot the collective mode spectrum of bilayer
graphene at two different densities, (a) $n=10^{13}cm^{-2}$ (high
density, $E_F > \omega_{SO}$) and (b) $n=10^{12}cm^{-2}$ (low density,
$E_F < \omega_{SO}$) in
Fig.~\ref{fig_blg_mode}. Here, $\omega_q$ is the uncoupled plasmon
frequency and the shaded regions represent the intraband and interband
particle-hole continuum. The corresponding loss functions are plotted
in Fig.~\ref{fig_spec_blg}.  

In the high density limit where $w_q \sim \omega_{SO}$, 
there is a strong plasmon-phonon coupling
as evidenced by the deviations of $\omega_+$ from $\omega_{SO}$ and of
$\omega_-$ from $\omega_q$, which gives rise to
the gapped mode $\omega_+$ having a linear dispersion with a positive slope
in the low $q$ limit. At larger $q$ values, it approaches the
uncoupled plasmon dispersion, as seen in
Fig.~\ref{fig_blg_mode}(a). 
%One big difference from MLG is that 
%due to the softening of the singularity at the intraband
%continuum~\cite{sensarma2010} arising from enhanced backscattering 
%both the coupled and uncoupled plasmons
%($\omega_-$ and $\omega_q$ respectively) merge into the continuum and
%gets overdamped. 
The $\omega_-$ mode merges into the continuum at a
critical wavevector $q_c$, which is much smaller than that of the
uncoupled mode indicating strong electron-phonon coupling. Furthermore, as
seen from Fig.~\ref{fig_spec_blg}(a), the phonon-like mode $\omega_+$
carries a much larger spectral weight than $\omega_-$.
% and hence will
%be the most prominent mode seen in experiments.

In the low density limit where $\omega_q < \omega_{SO}$, 
the plasmon-phonon coupling is weak
and the gapped mode $\omega_+$ is barely affected by the coupling.
% and has almost the same energy of uncoupled SO-phonon. 
In addition, the mode energy decreases linearly
in the long wave length limit, as seen in Fig.~\ref{fig_blg_mode}(b).
% has a linear dispersion at
%low $q$ with a small negative slope, i.e. the mode frequcncy comes down in
%frequency before rising again and merging with the intraband continuum
%and getting overdamped,
The
small negative slope is a consequence of the coupling of the phonon
mode to the interband
single particle excitation \cite{sensarma2010}
arising from the enhanced backscattering in the system and is a
distinct difference between the MLG and BLG. Note
that when the SO phonon
mode is pushed into the 
interband electron-hole continuum, 
the coupled $\omega_+$ mode is always Landau-damped due to the presence of the
interband continuum and carries little spectral weight beyond a very
small range of low $q$ values. The deviation of the plasmon-like mode
$\omega_-$ from the uncoupled dispersion is much smaller than in the
high density limit, further showing that the plasmon-phonon coupling
is weak in this limit. From Fig.~\ref{fig_spec_blg}(b), we find that
beyond a small range of low $q$ values, the plasmon-like mode carries
much more spectral weight than the phonon-like mode and hence, at low
densities, the plasmon mode should be easier to detect in BLG.

In summary, we have calculated the dispersion and the spectral weight of 
the coupled plasmon-phonon mode of 2D graphene.
We find that the mode-coupling effect 
is strong in monolayer graphene at all densities in contrast
to the corresponding bilayer graphene, where the coupling is only
significant at high densities.
Since the carriers in graphene are strongly coupled to the
surface optical phone of a polar substrate 
it is important to understand the many-body
renormalization of the single-particle properties
and the transport
properties in the presence of electron-SO phonon coupling.
% which has not been considered in most literature. 

This work is supported by ONR-MURI and SWAN-NRI.

%\bibliographystyle{apsrev}
%\bibliography{coupling}
%Merlin.mbs v4.21 2009-07-09.
%
\end{document}